\documentclass[twoside,12pt]{article}
\usepackage{indentfirst}
\usepackage{bm}
\usepackage{graphicx}
\usepackage{epsfig}
\usepackage{amsmath}
\usepackage{amsfonts}
\usepackage{amssymb}
\usepackage{amsthm}
\usepackage{latexsym}

\usepackage{booktabs}
\usepackage{caption}
\newtheorem{thm}{Theorem}

\usepackage{epsfig}
\usepackage{amsmath}
\usepackage{epstopdf}
\usepackage{pgf,fancyhdr}
\usepackage{float}
\usepackage{color,xcolor}
\usepackage{soul}
\soulregister{\cite}7 
\soulregister{\citep}7 
\soulregister{\citet}7 
\soulregister{\ref}7 
\soulregister{\pageref}7 
\soulregister{\textbf}7
\soulregister{\textit}7
\newcommand{\tr}{\mathrm{tr}}

\begin{document}

\begin{center}
\LARGE\bf Detection of tripartite entanglement based on principal basis matrix representations
\end{center}

\begin{center}
\rm  Hui Zhao,$^1$ \  Yu-Qiu Liu,$^1$ \ Shao-Ming Fei,$^{2, 3}$ \ Zhi-Xi Wang$^2$ \ and Naihuan Jing$^{4}$
\end{center}

\begin{center}
\begin{footnotesize} \sl
$^1$ Department of Mathematics, Faculty of Science, Beijing University of Technology, Beijing 100124, China

$^2$ School of Mathematical Sciences,  Capital Normal University,  Beijing 100048,  China

$^3$ Max-Planck-Institute for Mathematics in the Sciences, Leipzig 04103, Germany

$^4$ Department of Mathematics, North Carolina State University, Raleigh, NC 27695, USA

\end{footnotesize}
\end{center}

\vspace*{2mm}

\begin{center}
\begin{minipage}{15.5cm}
\parindent 20pt\footnotesize
We study the entanglement in tripartite quantum systems by using the principal basis matrix representations of density matrices. Using the Schmidt decomposition and local unitary transformation, we first convert the general states to simpler forms and then construct some special matrices from the correlation tensors of the simplified density matrices. Based on the different linear combinations of these matrices, necessary conditions are presented to detect entanglement of tripartite states. Detailed examples show that our method can detect more entangled states than previous ones.
\end{minipage}
\end{center}

\begin{center}
\begin{minipage}{15.5cm}
\begin{minipage}[t]{2.3cm}{\bf Keywords:}\end{minipage}
\begin{minipage}[t]{13.1cm}
Genuine multipartite entanglement, Principal basis, Correlation tensor
\end{minipage}\par\vglue8pt
{\bf PACS: }03.65.Ud, 02.10.Ox, 03.67.Mn
\end{minipage}
\end{center}

\section{Introduction}
Quantum entanglement is the key resource in many information processing tasks such as quantum cryptography \cite{ea}, teleportation \cite{bcb} and dense coding \cite{bch}. The genuine multipartite entanglement (GME) has particularly significant advantages such as in highly sensitive metrological tasks \cite{hpl,tg}. It is of importance to study proper descriptions and effective detections of the genuine multipartite entanglement for given quantum states.

A lot of researches have been done towards the detection of entanglement and genuine multipartite entanglement \cite{hgy,lgy,ghl}. The methods to detect k-nonseparability and k-partite entanglement by using of quantum Fisher information were proposed in \cite{hqg}. By considering the basis dependence of the permutation operation, the authors in \cite{gyv} obtained strong bounds on a multipartite nonseparability measure. The separability criteria to identify genuinely entangled and nonseparable n-partite mixed quantum states were derived in \cite{gh}. The results with experimentally implemented processes indicated that the compressive method works equally well with both entangled and product input states and output measurement resources even in the presence of noise \cite{tsk}. The authors in \cite{tgg} presented entanglement witness operators for detecting GME which require only two local measurement settings. A sufficient criterion for detecting GME was derived by using local sum uncertainty relations \cite{ljc}. L. M. Yang et. al. \cite{lmy} proposed a criterion for detecting genuine tripartite entanglement based on quantum Fisher information. Y. Akbari-Kourbolagh \cite{aky} found sufficient criteria for detecting the entanglement of three-qubit states in the vicinity of GHZ, W states and the PPT entangled states. Based on the Bloch representation of density matrices, several criteria to detect GME were also derived. In \cite{lmj} the authors gave a sufficient condition for GME and a lower bound for the GME concurrence by using the norms of correlation tensor. M. Li et. al. \cite{lmf} studied the separability of tripartite states by introducing an operator related to the transformations on the subsystem. In \cite{ssq} the authors uncovered some separability criteria of bipartite and multipartite entanglement.

In this paper, we study the entanglement of tripartite quantum systems by using the principal basis matrix presentation of density matrices \cite{lbgj}. Consider the $d$-dimensional Hilbert spaces $H^d$ with  orthonormal basis $\{|a\rangle\}_{a=0}^{d-1}$. Let $E_{ij}$ be the $d\times d$ matrix with the only nonzero entry $1$ at the position $(i,j)$, and $\omega$ be a fixed $d$-th primitive root of unity. The principal basis matrices are given by
$A_{ij}=\sum\limits_{m\in Z_{d}}\omega^{im}E_{m,m+j}$,
where $\omega^{d}=1$, $i,j\in Z_{d}$ and $Z_{d}$ is $Z$ modulo $d$. It is readily apparent that
$\tr(A_{ij}A_{kl}^{\dagger})=\delta_{ik}\delta_{jl}d $ \cite{hjz}.

We use the principal basis matrices to expand a given density matrix. Then we employ Schmidt decomposition \cite{man} and local unitary (LU) transformations \cite{JYZ} to transform the density matrix into a simpler form so as to construct useful invariants of GME.
For any bipartite state $\rho$ in $H_1^{d_1}\otimes H_2^{d_2}$, $\rho$ can be expressed as $\rho=\frac{1}{d_1d_2}(I_{d_1}\otimes I_{d_2}+\sum\limits_{(i,j)}a_{ij}A_{ij}^{(1)}\otimes I_{d_2}+\sum\limits_{(i,j)}b_{ij}I_{d_1}\otimes A_{ij}^{(2)}
+\sum\limits_{(i,j),(k,l)}c_{ij}^{kl}A_{ij}^{(1)}\otimes A_{kl}^{(2)})$.
Define the $(d_1^2-1)\times(d_2^2-1)$ matrix $S=[c_{ij}^{kl}]$. For example, when $\rho$ in $H_1^{2}\otimes H_2^{3}$, $$S=\left[\footnotesize
                                \begin{array}{cccccccc}
                                  c_{01}^{01} & c_{01}^{02} & c_{01}^{10} & c_{01}^{11} & c_{01}^{12} & c_{01}^{20} & c_{01}^{21} & c_{01}^{22} \\
                                  c_{10}^{01} & c_{10}^{02} & c_{10}^{10} & c_{10}^{11} & c_{10}^{12} & c_{10}^{20} & c_{10}^{21} & c_{10}^{22} \\
                                  c_{11}^{01} & c_{11}^{02} & c_{11}^{10} & c_{11}^{11} & c_{11}^{12} & c_{11}^{20} & c_{11}^{21} & c_{11}^{22} \\
                                \end{array}
                              \right].
$$
Denote $UAU^\dagger$ by $A^U$. Let $\rho'=\rho^{(U_1\otimes U_2)}$, where $U_1$ and $U_2$ are unitary matrices. One sees that $\sum\limits_{(i,j),(k,l)}c_{ij}^{kl}(A_{ij}^{(1)})^{U_1}\otimes (A_{kl}^{(2)})^{U_2}=\sum\limits_{(i,j),(k,l)}(\sum\limits_{(i',j'),(k',l')}m_{i'j'}^{ij}c_{i'j'}^{k'l'}n_{k'l'}^{kl})
A_{ij}^{(1)}\otimes A_{kl}^{(2)}$, where $M=[m_{ij}^{i'j'}]$ and $N=[n_{kl}^{k'l'}]$ are two unitary matrices. Therefore, $S(\rho')=M^tS(\rho)N$. Let $\|P\|_{tr}$ stand for the trace norm of a matrix $P$ defined by $\|P\|_{tr}=\sum_{i}\sigma_{i}=\tr\sqrt{P^{\dagger}P}$, $P\in\mathbb{R}^{m\times n}$, where $\sigma_{i}$ $(i=1,2,\cdot\cdot\cdot,min\{m,n\})$ are the singular values of the matrix $P$. Then we have $\|S(\rho')\|_{tr}=\|S(\rho)\|_{tr}$ due to the fact that the singular values of a rectangular matrix
$S$ are the same as those of $M^tSN$ when $M$ and $N$ are unitary matrices. Thus, the study of $\rho$ can be translated into that of $\rho'$.
Using this idea, we can simplify the matrices constructed by correlation tensors.

This paper is organized as follows. In Section 2, we derive the criteria to detect entanglement of three-qubit states. By detailed examples, our results are seen to outperform some previously available results. In Section 3, we generalize the results to tripartite qudit quantum systems. Conclusions are given in Section 4.

\section{Entanglement for three-qubit quantum states}
We first consider the entanglement for tripartite qubit states. Let $H_{i}^{d}$ denote the $i$-th $d$-dimensional Hilbert space. A tripartite pure state $|\varphi\rangle\in H_{1}^{d}\otimes H_{2}^{d}\otimes H_{3}^{d}$ is called biseparable under the bipartition $f|gh$ if $|\varphi\rangle$ can be written as $|\varphi\rangle=|\varphi_{f}\rangle\otimes|\varphi_{gh}\rangle$, otherwise it is called entanglement under the bipartition $f|gh$, where $|\varphi_{f}\rangle$ and $|\varphi_{gh}\rangle$ denote respectively the pure states in $H_{f}^{d}$ and $H_{g}^{d}\otimes H_{h}^{d}$ $(f\neq g\neq h\in\{1,2,3\})$.

For $d=2$, the principal basis matrices are given by
\begin{equation}\label{1}
  A_{00}=\left[
           \begin{array}{cc}
             1 & 0 \\
             0 & 1 \\
           \end{array}
         \right],~
  A_{01}=\left[
           \begin{array}{cc}
             0 & 1 \\
             1 & 0 \\
           \end{array}
         \right],~
  A_{10}=\left[
           \begin{array}{cc}
             1 & 0 \\
             0 & -1 \\
           \end{array}
         \right],~
  A_{11}=\left[
           \begin{array}{cc}
             0 & 1 \\
             -1 & 0 \\
           \end{array}
         \right].
\end{equation}
A general three-qubit state $\rho\in H_{1}^{2}\otimes H_{2}^{2}\otimes H_{3}^{2}$ can be expressed as,
\begin{equation}\label{2}
\begin{split}
\rho=&\frac{1}{8}(I_{2}\otimes I_{2}\otimes I_{2}+\sum_{(i,j)}u_{ij}A_{ij}\otimes I_{2}\otimes I_{2}+\sum_{(k,l)}v_{kl}I_{2}\otimes A_{kl}\otimes I_{2}
+\sum_{(s,t)}w_{st}I_{2}\otimes I_{2}\otimes A_{st}\\
&+\sum_{(i,j),(k,l)}x_{ij,kl}A_{ij}\otimes A_{kl}\otimes I_{2}+\sum_{(i,j),(s,t)}y_{ij,st}A_{ij}\otimes I_{2}\otimes A_{st}+\sum_{(k,l),(s,t)}z_{kl,st}I_{2}\otimes A_{kl}\otimes A_{st}\\
&+\sum_{(i,j),(k,l),\atop (s,t)}r_{ij,kl,st}A_{ij}\otimes A_{kl}\otimes A_{st}),
\end{split}
\end{equation}
where $I_{2}$ denotes the $2\time 2$ identity matrix, the summation indices in $(i,j)$ ($(k,l)$ and $(s,t)$) are not both zero, $r_{ij,kl,st}=\tr(\rho A_{ij}^{\dagger}\otimes A_{kl}^{\dagger}\otimes A_{st}^{\dagger})$. Denote by $T_{1}^{1|23}$, $T_{2}^{1|23}$, $T_{3}^{1|23}$, $T_{1}^{2|13}$, $T_{2}^{2|13}$, $T_{3}^{2|13}$, $T_{1}^{3|12}$, $T_{2}^{3|12}$ and $T_{3}^{3|12}$ the matrices with entries $r_{01,kl,st}$, $r_{10,kl,st}$, $r_{11,kl,st}$, $r_{ij,01,st}$, $r_{ij,10,st}$, $r_{ij,11,st}$, $r_{ij,kl,01}$, $r_{ij,kl,10}$ and $r_{ij,kl,11}$, respectively. For example,
$$T_1^{2|13}=\left[
    \begin{array}{ccc}
      r_{01,01,01} & r_{01,01,10} & r_{01,01,11} \\
      r_{10,01,01} & r_{10,01,10} & r_{10,01,11} \\
      r_{11,01,01} & r_{11,01,10} & r_{11,01,11} \\
    \end{array}
  \right],
$$
and the other matrices are arranged similarly. Set $S^{f|gh}=a_fT_1^{f|gh}+b_fT_2^{f|gh}+c_fT_3^{f|gh}$, where $a_f, b_f$ and $c_f$ are real constants, $f\neq g\neq h\in\{1,2,3\}$.

Let $\rho'=\rho^{(I\otimes U_2 \otimes U_3)}$, where $U_2, U_3\in U(2)$ are $2\times 2$ unitary matrices. Assume $A_{ij}^{U_{2}}=\sum\limits_{(i',j')\neq(0,0)}m_{ij,i'j'}A_{i'j'}$ and $A_{ij}^{U_{3}}=\sum\limits_{(i',j')\neq(0,0)}n_{ij,i'j'}A_{i'j'}$ with coefficient matrices $M=[m_{ij,i'j'}]$ and $N=[n_{ij,i'j'}]$. Then we have
\begin{equation}\label{3}
T_1^{1|23}(\rho')=M^tT_1^{1|23}(\rho)N,~~T_2^{1|23}(\rho')=M^tT_2^{1|23}(\rho)N,~~ T_3^{1|23}(\rho')=M^tT_3^{1|23}(\rho)N.
\end{equation}
Therefore, $S^{1|23}(\rho')=M^tS^{1|23}(\rho)N$. Let $S^{1|23}(\rho)=UDV$ be the singular value decomposition of $S^{1|23}(\rho)$, where $U$ and $V$ are unitary matrices, $D$ is a diagonal matrix given by the singular values of $S^{1|23}(\rho)$. Then $S^{1|23}(\rho')=M^tUDVN$ has the same singular values as $S^{1|23}(\rho)$. Therefore, $\|S^{1|23}(\rho')\|_{tr}=\|S^{1|23}(\rho)\|_{tr}$. Namely, the trace norm is invariant under local unitary transformations.

Let us now consider the biseparable pure states. If $\rho=|\varphi\rangle\langle\varphi|$ is separable under the bipartition $1|23$, i.e., $|\varphi\rangle=|\varphi_1\rangle\otimes|\varphi_{23}\rangle\in H_1^2\otimes H_{23}^4$, using the Schmidt decomposition one has $|\varphi\rangle=t_0|0\alpha\rangle+t_1|1\beta\rangle$ under some suitable local bases, where $t_0^2+t_1^2=1$. Using LU equivalence, when $|\varphi_{23}\rangle\in H_{23}^4$ is separable we can transform $\{|\alpha\rangle,|\beta\rangle\}$ into two orthonormal bases which constitute separable states: (i) $\{|\alpha\rangle,|\beta\rangle\}=\{|00\rangle,|01\rangle\}$; when $|\varphi_{23}\rangle\in H_{23}^4$ is entangled, we can transform $\{|\alpha\rangle,|\beta\rangle\}$ into two orthonormal bases which constitute entangled states: (ii) $\{|\alpha\rangle,|\beta\rangle\}=\{|00\rangle,|11\rangle\}.$
The matrices $T_{1}^{1|23}$, $T_{2}^{1|23}$ and $T_{3}^{1|23}$ are then given by
\begin{equation}\label{4}
(i):~T_{1}^{1|23}=\left[
                    \begin{array}{ccc}
                      0 & 0 & 0 \\
                      2t_0t_1 & 0 & 0 \\
                      0 & 0 & 0 \\
                    \end{array}
                  \right],~
    T_{2}^{1|23}=\left[
                    \begin{array}{ccc}
                      0 & 0 & 0 \\
                      0 & t_0^2+t_1^2 & 0 \\
                      0 & 0 & 0 \\
                    \end{array}
                  \right],~
T_{3}^{1|23}=\left[
                    \begin{array}{ccc}
                      0 & 0 & 0 \\
                      0 & 0 & 2t_0t_1 \\
                      0 & 0 & 0 \\
                    \end{array}
                  \right];
\end{equation}
\begin{equation}\label{5}
(ii):~T_{1}^{1|23}=\left[
                    \begin{array}{ccc}
                    2t_0t_1 & 0 & 0 \\
                      0 & 0 & 0 \\
                      0 & 0 & 2t_0t_1 \\
                    \end{array}
                  \right],~
    T_{2}^{1|23}=\left[
                    \begin{array}{ccc}
                      0 & 0 & 0 \\
                      0 & t_0^2-t_1^2 & 0 \\
                      0 & 0 & 0 \\
                    \end{array}
                  \right],~
    T_{3}^{1|23}=\left[
                    \begin{array}{ccc}
                      0 & 0 & 2t_0t_1 \\
                      0 & 0 & 0 \\
                      2t_0t_1 & 0 & 0 \\
                    \end{array}
                  \right],
\end{equation}
corresponding to the cases (i) and (ii) respectively.

\begin{thm} For a biseparable pure state $\rho$, we have\\
(1) If $\rho$ is separable under the bipartition $1|23$, then $\|S^{1|23}\|_{tr}=\sqrt{4(a_1^2+c_1^2)t_0^2t_1^2+b_1^2}$~~or\\$|b_1|\sqrt{(t_0^2-t_1^2)^2}+2\mu_1t_0t_1$;\\
(2) If $\rho$ is separable under the bipartition $2|13$, then $\|S^{2|13}\|_{tr}=|b_2|(1+4t_0t_1)$
~~or\\$|b_2|\sqrt{(t_0^2-t_1^2)^2}+2\mu_2t_0t_1$;\\
(3) If $\rho$ is separable under the bipartition $3|12$, then $\|S^{3|12}\|_{tr}=\sqrt{4(a_3^2+c_3^2)t_0^2t_1^2+b_3^2}$
~~or\\$|b_3|\sqrt{(t_0^2-t_1^2)^2}+2\mu_3t_0t_1$\\
corresponding to cases (i) or (ii), respectively, where $\mu_f=|a_f+c_f|+|a_f-c_f|$.

\begin{proof}
(1) By using the Schmidt decomposition and the expressions (4) and (5), the conclusion is easily seen.

(2) If a pure state $\rho=|\varphi\rangle\langle\varphi|$ is separable under $2|13$ bipartition, then $|\varphi\rangle=|\varphi_2\rangle\otimes|\varphi_{13}\rangle\in H_2^2\otimes H_{13}^4$.
Similarly we need to consider the following two cases:

The first case: $|\varphi_{13}\rangle$ is separable. Under LU we have $|\varphi\rangle=t_0|000\rangle+t_1|101\rangle$, and
\begin{align}\label{6}
~T_{1}^{2|13}=\textbf{0},~
    T_{2}^{2|13}=\left[
                    \begin{array}{ccc}
                      2t_0t_1 & 0 & 0 \\
                      0 & t_0^2+t_1^2 & 0 \\
                      0 & 0 & 2t_0t_1 \\
                    \end{array}
                  \right],~
    T_{3}^{2|13}=\textbf{0},
\end{align}
where \textbf{0} denotes the zero matrix. Then $\|S^{2|13}\|_{tr}=|b_2|\cdot[(t_0^2+t_1^2)+4t_0t_1]=|b_2|(1+4t_0t_1).$

The second case: $|\varphi_{13}\rangle$ is entangled. Under LU we have $|\varphi\rangle=t_0|000\rangle+t_1|111\rangle$, and
\begin{align}\label{7}
T_{1}^{2|13}=\left[
                    \begin{array}{ccc}
                    2t_0t_1 & 0 & 0 \\
                      0 & 0 & 0 \\
                      0 & 0 & 2t_0t_1 \\
                    \end{array}
                  \right],~
T_{2}^{2|13}=\left[
                    \begin{array}{ccc}
                      0 & 0 & 0 \\
                      0 & t_0^2-t_1^2 & 0 \\
                      0 & 0 & 0 \\
                    \end{array}
                  \right],~
T_{3}^{2|13}=\left[
                    \begin{array}{ccc}
                      0 & 0 & 2t_0t_1 \\
                      0 & 0 & 0 \\
                      2t_0t_1 & 0 & 0 \\
                    \end{array}
                  \right].
\end{align}
We obtain $\|S^{2|13}\|_{tr}=|b_2|\sqrt{(t_0^2-t_1^2)^2}+2\mu_2t_0t_1$.

(3) Since $S^{3|12}=(S^{1|23})^t$ for the case (i) and $S^{3|12}=S^{1|23}$ for the case (ii), the conclusion can be proved in a similar way.
\end{proof}
\end{thm}

{\bf Remark} There are two kinds of genuinely three-qubit entangled pure states under stochastic local operations and classical communication (SLOCC) \cite{wgj}, namely, the $GHZ$ state and $W$ state. We mix the GHZ state or W state with the white noise, by choosing different values of $\{a_f, b_f, c_f\}$, accordingly, we can obtain different upper bounds of $\|S^{f|gh}\|_{tr}$ to detect entanglement.

For mixed states, we have the following corollaries.

\textbf{Corollary 1} For a quantum mixed state $\rho$, if $\|S^{2|13}\|_{tr}>3|b_2|$, where $3|b_2|<\mu_2$ $(b_2, \mu_2\neq0)$, then $\rho$ is entangled.

\textbf{Corollary 2} For a quantum mixed state $\rho$, if $\|S^{f|gh}\|_{tr}=\|a_fT_1^{f|gh}+b_fT_2^{f|gh}\|_{tr}>\sqrt{b_f^2+4a_f^2}$, where $|\frac{a_f}{b_f}|<1.6248$, then $\rho$ is entangled.

The proofs of the above corollaries can be seen respectively from the analysis of the following examples.

\textbf{Example 1} Consider the mixed state $\rho_{GHZ}$,
\begin{equation}\label{8}
\rho_{GHZ}=\frac{x}{8}I_8+(1-x)|GHZ\rangle\langle GHZ|, 0\leq x\leq1,
\end{equation}
where $|GHZ\rangle=\frac{1}{\sqrt{2}}(|000\rangle+|111\rangle)$.
We have $\|S^{2|13}(\rho_{GHZ})\|_{tr}=\mu_2(1-x)$.
By using Lemma 1, when a mixed state $\rho=\sum p_i\rho_i\in H_1^2\otimes H_2^2\otimes H_3^2$ $(0<p_i\leq1$, $\sum p_i=1)$ is biseparable, we have
\begin{equation}\label{9}
\|S^{2|13}(\rho)\|_{tr}\leq\sum p_i\|S^{2|13}(\rho_i)\|_{tr}\leq|b_2|(1+4t_0t_1)\leq3|b_2|.
\end{equation}
Therefore, if $\rho_{GHZ}$ is separable under the bipartition $2|13$, we have $x\geq1-\frac{3|b_2|}{\mu_2}$. Consequently, if $x<1-\frac{3|b_2|}{\mu_2}$, $\rho_{GHZ}$ is an entangled state. Since $x\in[0,1]$, we can detect the entanglement for $\rho_{GHZ}$ when $3|b_2|<\mu_2$.
By using Corollary 1, we get Table 1 of the entanglement ranges for $\rho_{GHZ}$. Obviously, these results are better than $0\leq x<\frac{2}{3}\approx0.6667$ given in \cite{aky}.
\begin{table}
\footnotesize
\centering
\begin{tabular}{|c|c|}
  \hline
  \  & the range of entanglement \\
  \hline
  $a_2=-4, b_2=1, c_2=6$ & $0\leq x<0.75$ \\
  \hline
  $a_2=3, b_2=1, c_2=7$ & $0\leq x<0.7857$ \\
  \hline
  $a_2=5, b_2=\frac{1}{3}, c_2=5$ & $0\leq x<0.9$ \\
  \hline
\end{tabular}
  \caption{The entanglement ranges for $\rho_{GHZ}$ with respect to the the corresponding coefficients.}\label{2}
\end{table}

\textbf{Example 2} Consider the mixed state $\rho_W$,
\begin{equation}\label{10}
  \rho_W=\frac{1-x}{8}I_8+x|W\rangle\langle W|, 0\leq x\leq1,
\end{equation}
where $|W\rangle=\frac{1}{\sqrt{3}}(|100\rangle+|010\rangle+|001\rangle)$. When $c_f=0$ we have
\begin{equation}\label{11}
\begin{split}
\|S^{f|gh}(\rho_W)\|_{tr}=&\|a_fT_1^{f|gh}+b_fT_2^{f|gh}\|_{tr}\\
=&\frac{1}{\sqrt{18}}(\sqrt{8}+\sqrt{13+8(\frac{a_f}{b_f})^2+\sqrt{25+16(\frac{a_f}{b_f})^2}}\\
&+\sqrt{13+8(\frac{a_f}{b_f})^2-\sqrt{25+16(\frac{a_f}{b_f})^2}})|b_f|x.
\end{split}
\end{equation}
By using Lemma 1, when a mixed state $\rho=\sum p_i\rho_i\in H_1^2\otimes H_2^2\otimes H_3^2$ $(0<p_i\leq1$, $\sum p_i=1)$ is biseparable, we have
\begin{equation}\label{12}
\begin{split}
\|S^{f|gh}(\rho)\|_{tr}&\leq\sum p_i\|S^{f|gh}(\rho_i)\|_{tr}
\leq|b_f|\sqrt{(1-2t_1^2)^2}+2\mu_ft_1\sqrt{1-t_1^2}
\leq\sqrt{b_f^2+\mu_f^2},
\end{split}
\end{equation}
where the upper bound is obtained by taking the extreme value of the function with independent variable $t_1$. If $\rho_W$ is separable under the bipartition $f|gh$, we have $\|S^{f|gh}(\rho_W)\|_{tr}\leq\sqrt{b_f^2+4a_f^2}$. Consequently, if
\begin{equation}\label{13}
x>\frac{\sqrt{18}\sqrt{1+4(\frac{a_f}{b_f})^2}}{\sqrt{8}
+\sqrt{13+8(\frac{a_f}{b_f})^2+\sqrt{25+16(\frac{a_f}{b_f})^2}}
+\sqrt{13+8(\frac{a_f}{b_f})^2-\sqrt{25+16(\frac{a_f}{b_f})^2}}},
\end{equation}
$\rho_W$ is an entangled state. The function $f(\delta)$ of $\delta=\frac{a_f}{b_f}$ at the right end of (\ref{13}) is shown in Figure 1. Since $x\in[0,1]$, we can detect the entanglement for $\rho_W$ when $f(\delta)<1$, i.e., $|\delta|=|\frac{a_f}{b_f}|<1.6248$.
\begin{figure}[!htb]
  \centering
  \includegraphics[width=10
  cm]{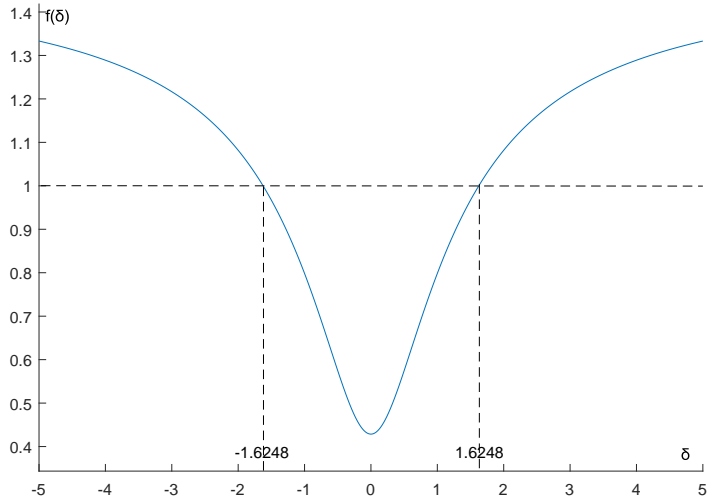}\\
  \caption{The image of function $f(\delta)$.}
\end{figure}

By using Corollary 2, we have Table 2 of entanglement ranges for $\rho_W$. These results are better than $0.6<x\leq1$ that was given in \cite{aae}. Moreover, when $a_f=0$ and $b_f=1$, our criterion can detect more entanglement than $0.4334<x\leq1$ which was given in \cite{yam}.
\begin{table}
\footnotesize
\centering
\begin{tabular}{|c|c|}
  \hline
  \  & the range of entanglement \\
  \hline
  $a_f=1, b_f=3$ & $0.5025<x\leq1$ \\
  \hline
  $a_f=1, b_f=10$ & $0.4361<x\leq1$ \\
  \hline
  $a_f=0, b_f=1$ & $0.4286<x\leq1$ \\
  \hline
\end{tabular}
  \caption{The entanglement ranges for $\rho_{W}$ with respect to the corresponding coefficients.}
\end{table}

\section{Entanglement for tripartite qudit quantum states}
Next, we consider entanglement of $(d\times d\times d)$-dimensional quantum states.
Different from the case of $d=2$ where all the principal basis matrices given in (\ref{1}) are real, for $d\geq 3$ the principal basis matrices are complex. For instance, for $d=3$ the principal basis matrices are given by,
$$\footnotesize
A_{01}=\left[
           \begin{array}{ccc}
             0 & 1 & 0 \\
             0 & 0 & 1 \\
             1 & 0 & 0 \\
           \end{array}
         \right],~
A_{02}=\left[
           \begin{array}{ccc}
             0 & 0 & 1 \\
             1 & 0 & 0 \\
             0 & 1 & 0 \\
           \end{array}
         \right],~
A_{10}=\left[
           \begin{array}{ccc}
             1 & 0 & 0 \\
             0 & \omega & 0 \\
             0 & 0 & \omega^2 \\
           \end{array}
         \right],~
A_{11}=\left[
           \begin{array}{ccc}
             0 & 1 & 0 \\
             0 & 0 & \omega \\
             \omega^2 & 0 & 0 \\
           \end{array}
         \right],$$
$$\footnotesize
A_{12}=\left[
           \begin{array}{ccc}
             0 & 0 & 1 \\
             \omega & 0 & 0 \\
             0 & \omega^2 & 0 \\
           \end{array}
         \right],~
A_{20}=\left[
           \begin{array}{ccc}
             1 & 0 & 0 \\
             0 & \omega^2 & 0 \\
             0 & 0 & \omega \\
           \end{array}
         \right],~
A_{21}=\left[
           \begin{array}{ccc}
             0 & 1 & 0 \\
             0 & 0 & \omega^2 \\
             \omega & 0 & 0 \\
           \end{array}
         \right],~
A_{22}=\left[
           \begin{array}{ccc}
             0 & 0 & 1 \\
             \omega^2 & 0 & 0 \\
             0 & \omega & 0 \\
           \end{array}
         \right]$$
and $A_{00}=I_3$, where $\omega^3=1$. 

For any state $\rho\in H_1^d\otimes H_2^d\otimes H_3^d$ $(d\geq3)$, $\rho$ has the principal basis representation:
\begin{equation*}
\begin{split}
\rho=&\frac{1}{d^3}(I_d\otimes I_d\otimes I_d+\sum_{(i,j)} u_{ij}A_{ij}\otimes I_{d}\otimes I_{d}+\sum_{(k,l)}v_{kl}I_{d}\otimes A_{kl}\otimes I_{d}+\sum_{(s,t)}w_{st}I_{d}\otimes I_{d}\otimes A_{st}\\
&+\sum_{(i,j),(k,l)}x_{ij,kl}A_{ij}\otimes A_{kl}\otimes I_{d}+\sum_{(i,j),(s,t)}y_{ij,st}A_{ij}\otimes I_{d}\otimes A_{st}+
\sum_{(k,l),(s,t)}z_{kl,st}I_{d}\otimes A_{kl}\otimes A_{st}\\
&+\sum_{(i,j),(k,l),\atop(s,t)}r_{ij,kl,st}A_{ij}\otimes A_{kl}\otimes A_{st}),
\end{split}
\end{equation*}
where the summation indices in $(i,j)$, $(k,l)$, $(s,t)$ are not both simultaneously zero, $r_{ij,kl,st}=tr(\rho A_{ij}^{\dagger}\otimes A_{kl}^{\dagger}\otimes A_{st}^{\dagger})$. Denote by $T_{i+1}^{1|23}$, $T_{k+1}^{2|13}$ and $T_{s+1}^{3|12}$ the matrices with entries $r_{i1,kl,st}$, $r_{ij,k1,st}$ and $r_{ij,kl,s1}$, respectively, for example, when $d=4$,
$$T_3^{1|23}=\left[\footnotesize
     \begin{array}{cccccc}
       r_{21,01,01} & r_{21,01,02} & r_{21,01,03} & r_{21,01,10} & \cdots & r_{21,01,33} \\
       r_{21,02,01} & r_{21,02,02} & r_{21,02,03} & r_{21,02,10} & \cdots & r_{21,02,33} \\
       r_{21,03,01} & r_{21,03,02} & \cdot & \cdot & \cdots & \cdot \\
       r_{21,10,01} & r_{21,10,02} & \cdot & \cdot & \cdots & \cdot \\
       \vdots & \vdots & \vdots & \vdots & \  & \vdots \\
       r_{21,13,01} & r_{21,13,02} & \cdot & \cdot & \cdots & \cdot \\
       \vdots & \vdots & \vdots & \vdots & \vdots & \vdots \\
       r_{21,30,01} & r_{21,30,02} & \cdot & \cdot & \cdots & \cdot \\
       \vdots & \vdots & \vdots & \vdots & \  & \vdots \\
       r_{21,33,01} & r_{21,33,02} & \cdot & \cdot & \cdots & \cdot \\
     \end{array}
   \right],$$
the other matrices are arranged in a similar way. Set $N^{1|23}=\sum\limits_iT_{i+1}^{1|23}$, $N^{2|13}=\sum\limits_k\omega^kT_{k+1}^{2|13}$, $N^{3|12}=\sum\limits_s\omega^sT_{s+1}^{3|12}$ and $T(\rho)=\frac{1}{3}(\|N^{1|23}\|_{tr}+\|N^{2|13}\|_{tr}+\|N^{3|12}\|_{tr})$, where $i,j,k,l,s,t\in Z_d$.

\begin{thm} For a pure state $\rho\in H_1^d\otimes H_2^d\otimes H_3^d$ $(d\geq3)$, we have\\
(1) If $\rho$ is separable under bipartition $1|23$, then $\|N^{1|23}\|_{tr}\leq\frac{\sqrt{d^3(d-1)}}{2}~~or~~ \frac{d^2}{2}$;\\
(2) If $\rho$ is separable under bipartition $2|13$, then $\|N^{2|13}\|_{tr}\leq\frac{d^2}{2}$;\\
(3) If $\rho$ is separable under bipartition $3|12$, then $\|N^{3|12}\|_{tr}\leq\frac{\sqrt{d^3(d-1)}}{2}~~or~~ \frac{d^2}{2}$.

\begin{proof}
(1) If $\rho$ is separable under bipartition $1|23$, we have
$|\varphi_{1|23}\rangle=|\varphi_{1}\rangle\otimes|\varphi_{23}\rangle\in H_{1}^{d}\otimes H_{23}^{d^2}$,
where $H_{23}^{d^2}=H_2^d\otimes H_3^d$. It follows from the Schmidt decomposition that $|\varphi_{1|23}\rangle=t_{0}|0\alpha_0\rangle+t_{1}|1\alpha_1\rangle\cdots +t_{d-1}|d-1,\alpha_{d-1}\rangle$, where $\sum t_{i}^2=1$. Taking into account local unitary equivalence in $H_2^d\otimes H_3^d$, if $|\varphi_{23}\rangle\in H_{23}^{d^2}$ is separable, we can transform $\{|\alpha_0\rangle, |\alpha_1\rangle,\cdots, |\alpha_{d-1}\rangle\}$ into orthonormal bases which constitute separable states: (i) $\{|\alpha_0\rangle, |\alpha_1\rangle,\cdots, |\alpha_{d-1}\rangle\}=\{|00\rangle, |01\rangle,\cdots,|0,d-1\rangle\}$, i.e., $|\varphi_{1|23}\rangle=|\varphi_{1}\rangle\otimes|\varphi_{2}\rangle\otimes|\varphi_{3}\rangle$ is fully separable. If $|\varphi_{23}\rangle\in H_{23}^{d^2}$ is entangled, we can transform $\{|\alpha_0\rangle, |\alpha_1\rangle,\cdots, |\alpha_{d-1}\rangle\}$ into orthonormal bases which constitute entangled states: (ii) $\{|\alpha_0\rangle, |\alpha_1\rangle,\cdots, |\alpha_{d-1}\rangle\}=\{|00\rangle, |11\rangle,\cdots,|d-1,d-1\rangle\}$.

For the first case (i), $|\varphi\rangle=|000\rangle+|101\rangle+\cdots+|d-1,0,d-1\rangle$, we have $N^{1|23}=dt_0t_1|\gamma\rangle\langle\eta|$, where $|\gamma\rangle=\left[
\begin{array}{ccccccccc}
0 & \cdots & 1_{(1,d)} & \cdots & 1_{(1,2d)} & \cdots & 1_{(1,(d-1)d)} & \cdots & 0 \\
\end{array}
\right]^t$, and
$|\eta\rangle=\left[
                 \begin{array}{ccccccc}
                   1_{(1,1)} & \cdots & 1_{(1,d+1)} & \cdots & 1_{(1,(d-1)d+1)} & \cdots & 0 \\
                 \end{array}
               \right]^t$, $a_{(i,j)}$ denotes that the element $a$ is located at the $i$th row and $j$th column of the matrix.
Therefore,
$\|N^{1|23}\|_{tr}=dt_0t_1\sqrt{d(d-1)}\leq\frac{\sqrt{d^3(d-1)}}{2}$,
where we have used $\||\gamma\rangle\langle\eta|\|_{tr}=\||\gamma\rangle\|\||\eta\rangle\|$ for vectors $|\gamma\rangle$ and $|\eta\rangle$. For the second case (ii), $|\varphi\rangle=|000\rangle+|111\rangle+\cdots+|d-1,d-1,d-1\rangle$, we have $N^{1|23}=dt_0t_1|\eta\rangle\langle\eta|$. Therefore,
$\|N^{1|23}\|_{tr}=dt_0t_1\sqrt{d^2}\leq\frac{d^2}{2}$.

(2) If $\rho$ is separable under bipartition $2|13$,
we need to consider again two cases. For the first case, $|\varphi\rangle=|000\rangle+|101\rangle+\cdots+|d-1,0,d-1\rangle$, and we have $T_{k+1}^{2|13}=\textbf{0}$ $(k\in Z_d)$, where \textbf{0} is the zero matrix. Therefore, $\|N^{2|13}\|_{tr}=0$. For the second case, $|\varphi\rangle=|000\rangle+|111\rangle+\cdots+|d-1,d-1,d-1\rangle$, one has
$N^{2|13}=dt_1t_2|\xi\rangle\langle\xi|,$
where $|\xi\rangle=\left[
                     \begin{array}{ccccccc}
                       1_{(1,1)} &\cdots & \overline{\omega}_{(1,d+1)}&\cdots\cdots&\overline{\omega^{d-1}}_{(1,(d-1)d+1)}&\cdots &0\\
                     \end{array}
                   \right]^t$, where $\overline{\omega}$ denotes the complex conjugate of $\omega$.
Then we have $\|N^{2|13}\|_{tr}=d^2t_1t_2\leq\frac{d^2}{2}$. 

(3) If $\rho$ is separable under bipartition $3|12$, for the first case,
$N^{3|12}=dt_1t_2|\xi\rangle\langle\gamma|$.
Then we have $\|N^{3|12}\|_{tr}=dt_1t_2\sqrt{d(d-1)}\leq\frac{\sqrt{d^3(d-1)}}{2}$.
For the second case, $N^{3|12}=dt_1t_2|\xi\rangle\langle\xi|$, we have
$\|N^{3|12}\|_{tr}=d^2t_1t_2\leq\frac{d^2}{2}$.
\end{proof}
\end{thm}


\section{Conclusion}
We have studied the entanglement in tripartite quantum systems. Based on the principal matrix representation, we constructed special matrices from the correlation tensors of the tripartite qubit states and have shown that they are invariant under local unitary transformation and are
capable to detect quantum entanglement.
From different linear combinations of these matrices, we have obtained new separability criteria and conditions on tripartite entanglement.
It has been shown that our criteria are more effective than some existing ones. By employing different methods for high dimensional cases, we have also studied the entanglement in $(d\times d\times d)$-dimensional quantum systems.

\textbf {Acknowledgements}
This work is supported by the National Natural Science Foundation of China under grant nos. 11101017, 11531004, 11726016, 12075159, 12126351 and 12171044,
Simons Foundation under grant no. 523868, Beijing Natural Science Foundation (Z190005), Academy for Multidisciplinary Studies, Capital Normal University, the Academician Innovation Platform of Hainan Province, and Shenzhen Institute for Quantum Science and Engineering, Southern University of Science and Technology (no. SIQSE202001).


\begin{thebibliography}{99}
\itemsep=-4pt plus.2pt minus.2pt
\small
\bibitem{ea} Ekert, A.K.: {\it Quantum cryptography based on Bell's theorem}. Phys. Rev. Lett. \textbf{67}, 661 (1991)
\bibitem{bcb} Bennett, C.H., Brassard, G., Jozsa, R. et al: {\it Teleporting an unknown quantum state via dual classical and Einstein-Podolsky-Rosen channels}. Phys. Rev. Lett. \textbf{70}, 1895 (1993)
\bibitem{bch} Bennett, C.H., Wiesner, S.J.: {\it Communication via One- and Two-Particle operators on Einstein-Podolsky-Rosen states}. Phys. Rev. Lett. \textbf{69}, 2881 (1992)
\bibitem{hpl} Hyllus, P., Laskowski, W., Krischek, R. et al: {\it Fisher information and multiparticle entanglement}. Phys. Rev. A \textbf{85}, 022321 (2012)
\bibitem{tg} T\'{o}th, G.: {\it Multipartite entanglement and high precision metrology}. Phys. Rev. A \textbf{85}, 022322 (2012)
\bibitem{hgy} Hong, Y., Gao, T., Yan, F. L.: {\it Detection of k-partite entanglement and k-nonseparability of multipartite quantum states.} Phys. Lett. A \textbf{401}, 127347 (2021)
\bibitem{lgy} Liu, L., Gao, T., Yan, F.: {\it Separability criteria via some classes of measurements.} Sci. China Phys. Mech. Astron. \textbf{60}, 100311 (2017)
\bibitem{ghl} Gao T., Hong Y., Lu, Y. et al: {\it Efficient $k$-separability criteria for mixed multipartite quantum states.} Europhysics Letters \textbf{104}, 20007 (2013)
\bibitem{hqg} Hong, Y., Qi, X. F., Gao, T. et al: {\it Detection of multipartite entanglement via quantum Fisher information.} Europhysics Letters \textbf{134}, 60006 (2021)
\bibitem{gyv} Gao, T., Yan, F. L., Van Enk S. J.: {\it Permutationally invariant part of a density matrix and nonseparability of n-qubit states.} Phys. Rev. Lett. \textbf{112}, 180501 (2014)
\bibitem{gh} Gao T., Hong Y.: {\it Detection of genuinely entangled and nonseparable $n$-partite quantum states.} Phys. Rev. A \textbf{82}, 062113 (2010)
\bibitem{tsk} Teo, Y. S., Struchalin, G. I., Kovlakov, E. V. et al: {\it Objective compressive quantum process tomography.} Phys. Rev. A \textbf{101}, 022334 (2020)
\bibitem{tgg} T\'{o}th G., G\"{u}hne O.: {\it Detecting genuine multipartite entanglement with two local measurements}. Phys. Rev. Lett. \textbf{94}, 060501 (2005)
\bibitem{ljc} Li, J., Chen, L.: {\it Detection of genuine multipartite entanglement based on uncertainty relations}. Quantum Inf. Process. \textbf{20}, 220 (2021)
\bibitem{lmy} Yang, L. M., Sun, B. Z., Chen, B. et al: {\it Quantum Fisher information-based detection of genuine tripartite entanglement}. Quantum Inf. Process. \textbf{19}, 262 (2020)
\bibitem{aky} Akbari-kourbolagh, Y.: {\it Entanglement criteria for the three-qubit states}. Int. J. Quantum Inf. \textbf{15}, 1750049 (2017)
\bibitem{lmj} Li, M., Jia, L.X., Wang, J. et al: {\it Measure and detection of genuine multipartite entanglement for tripartite systems}. Phys. Rev. A  \textbf{96}, 052314 (2017)
\bibitem{lmf} Li, M., Fei, S. M., Wang, Z. X.: {\it Separability of tripartite quantum systems}. Int. J. Quantum Inf.  \textbf{6}, 859 (2008)
\bibitem{ssq} Shen, S. Q., Yu, J., Li, M.:{\it Improved separability criteria based on Bloch representation of density matrices}. Sci. Rep.  \textbf{6}, 28850 (2016)
\bibitem{lbgj} Liu, M., Bai, C.~M., Ge, M.-L., Jing, N.: {\it Generalized Bell states and principal realization of the Yangian $Y(\mathfrak{sl}(N)$ }. J. Math. Phys. \textbf{54}, 021701,  (2013). 
\bibitem{hjz} Huang, X. F., Jing, N., Zhang, T. G.: {\it An upper bound of fully entangled fraction of mixed states}. Commun. in Theor. Phys.  \textbf{65}, 701 (2016)
\bibitem{man} Nielsen, M. A., Chuang, I. L.: {\it Quantum computation and quantum information}. Cambridge Univ. Press, Cambridge, (2000)
\bibitem{aae} Ac\'{i}n, A., Bru{\ss}, D., Lewenstein, M., Sanpera, A.: {\it Classification of mixed three-qubit states}. Phys. Rev. Lett. \textbf{87}, 040401 (2001)
\bibitem{yam} Akbari-Kourbolagh, Y.,  Azhdargalam, M.: {\it Entanglement criterion for tripartite systems based on local sum uncertainty relations}. Phys. Rev. Lett. \textbf{97}, 042333 (2018)
\bibitem{JYZ} Jing, N., Yang, M., Zhao, H.:{\it Local unitary equivalence of quantum states and simultaneous orthogonal equivalence}. J. Math. Phys. \textbf{57}, 062205 (2016)
\bibitem{wgj} D\"{u}r, W., Vidal, G., Cirac, J. I.:{\it Three qubits can be entangled in two inequivalent ways}. Phys. Rev. A  \textbf{62}, 062314 (2000)
\end{thebibliography}

\end{document}